\documentclass[aps,prl,twocolumn,epsfig,groupedaddress,showpacs]{revtex4}
\usepackage{latexsym}
\usepackage{epsfig}
\usepackage{graphicx}
\usepackage{textcomp}
\begin{document}

\title{Observation of ground-state quantum beats in atomic spontaneous emission}
\author{D. G. Norris${}^{1}$,
L. A. Orozco${}^{1}$, P. Barberis-Blostein${}^{1,2}$, H. J. Carmichael${}^{1,3}$}
\address{${}^{1}$Joint Quantum Institute, Department of Physics, University
of Maryland and\\ National Institute of Standards and Technology, College Park, MD 20742-4111, U.S.A.\\
${}^{2}$Instituto de Investigaci{\'o}n en Matem{\'a}ticas Aplicadas y en Sistemas, Universidad Nacional Aut{\'o}noma de M{\'e}xico, M{\'e}xico, DF 01000, M{\'e}xico\\${}^{3}$Department of Physics, University of Auckland, Private Bag 92019, Auckland, New Zealand.}

\date{\today}

\begin{abstract}
We report ground-state quantum beats in spontaneous emission from a continuously driven atomic ensemble.  Beats are visible only in an intensity autocorrelation and evidence spontaneously generated coherence in radiative decay.  Our measurement realizes a quantum eraser where a first photon detection prepares a superposition and a second erases the ``which-path" information in the intermediate state.


\end{abstract}


\maketitle

Quantum beats are oscillations in the radiation intensity of an ensemble of excited atoms due to interfering emission pathways. They must be counted amongst the earliest predictions of quantum mechanics \cite{breit33}.  
So-called ``Type-I'' atoms exhibit beats at the separation frequency of two excited states which, prepared in a superposition, decay to a common ground state.  The preparation may be achieved in a number of ways, \textit{e.g.},  through pulsed optical excitation \cite{dodd67} or cascade emission \cite{aspect84}. In all cases the coherence lasts for the excited state lifetime, unlike ground-state coherence, which can last long enough to be interrogated later and, for this reason, is favored by the field of quantum information.

We consider an unusual situation where \textit{ground} state coherence gives rise to a long-lived quantum beat in spontaneous emission. As repeatedly noted \cite{breit33,chow75,herman75}, QED predicts no beat in the decay of a ``Type-II'' atom to non-degenerate ground states, since they are orthogonal.  Nevertheless, while beats may be absent from the mean intensity, they can still lie hidden in the fluctuations.  In an elegant experiment in the 1950's Forrester \textit{et al.} \cite{forrester55} showed this for the (classical) beating of light from a pair of incoherent sources. We proceed in similar spirit; we recover a long-lived quantum beat from fluctuations.

In contrast to recent experiments \cite{wilk07b,weber09}, which aim for deterministic quantum control, the ground-state coherence in our experiment is both prepared and read out by spontaneous emission. Moreover, our measured beat is different from that seen by Schubert \textit{et al.} \cite{schubert95}, where an interference occurs in absorption rather than emission.


Creation of coherence through spontaneous emission, so-called spontaneously generated coherence, has been discussed in the theoretical literature \cite{javanainen92,patnaik99,economou05} and indirect experimental evidence reported for spontaneous creation of electron spin coherence in charged GaAs quantum dots \cite{dutt05}. We detect only spontaneous emission, and therefore make a direct and unambiguous observation of spontaneously generated coherence. We realize, in a continuously driven variation, the scheme of  Zajonc \cite{zajonc83} for generating ground-state quantum beats on the principle of the quantum eraser \cite{scully82}.

\begin{figure}[b]
\includegraphics[width=3.1in]{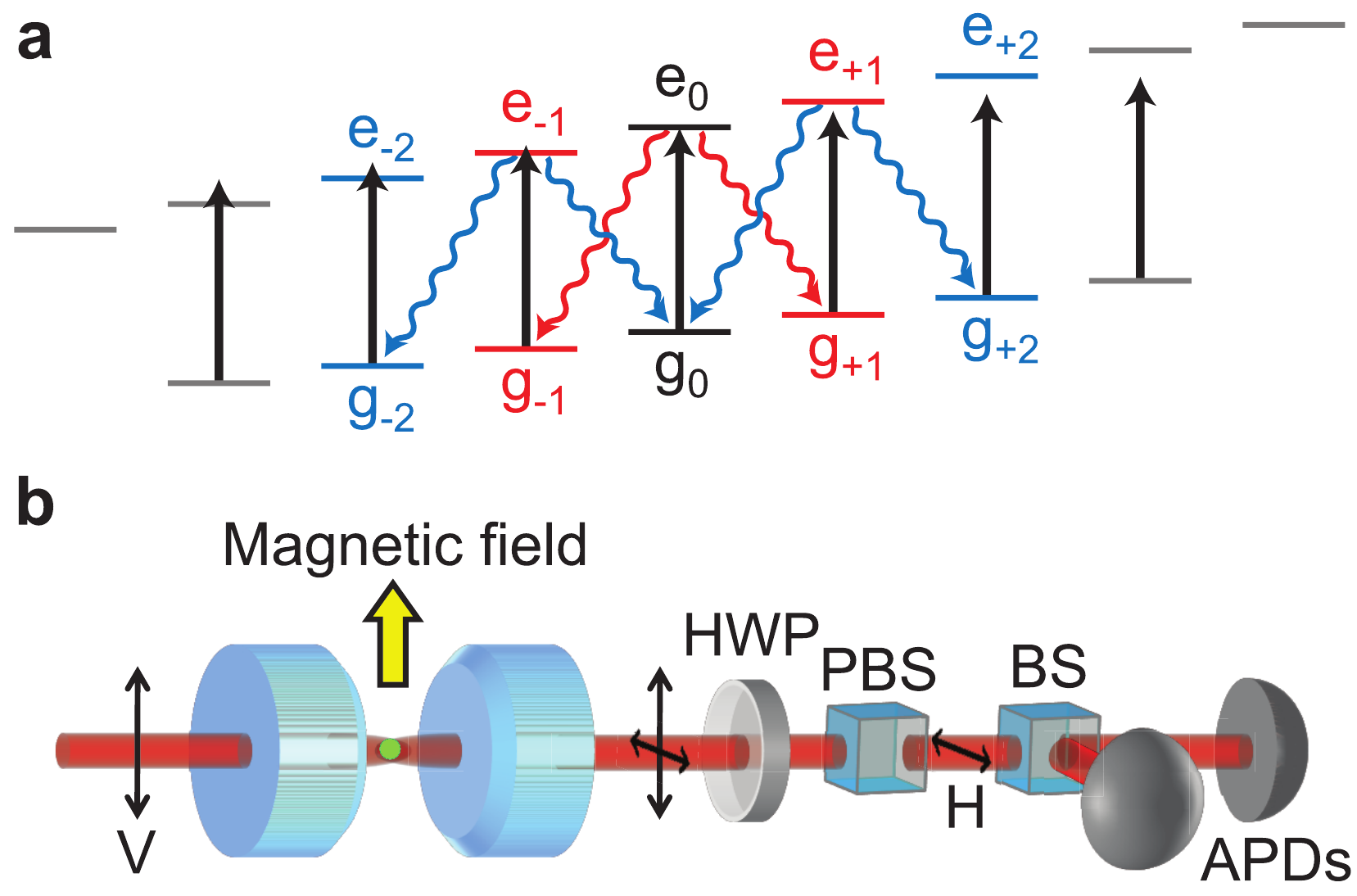}
\caption{\label{levels} (a) $\pi$-excitation of an $F=3$ to $F^\prime=4$ transition with scattering of a first (red) and second (blue) photon into the $H$ mode; b) schematic of the apparatus,  HWP: half-wave plate,  PBS: polarizing beam-splitter, BS: beam-splitter, APD: avalanche photodiode.}
\end{figure}

Beginning with an outline of the scheme, we consider first one idealized atom then a realistic atomic ensemble.
Consider an atom with Zeeman structure in its ground and excited states interacting with degenerate, orthogonally polarized cavity modes, $H$ and $V$; a weak magnetic field sets the quantization axis in the $V$ direction, and mode $V$ is weakly and continuously driven (Fig.~\ref{levels}).
The atom is prepared in state $|g_0\rangle$ from which it is excited to $|e_0\rangle$ by the $V$ mode. It may return to the ground state emitting a $\pi, \sigma^{+}$ or $\sigma^{-}$ photon, or any linear combination conserving angular momentum. In the given geometry, only $\sigma^{+}$ or $\sigma^{-}$ light couples to the $H$ mode, with the helicity undetermined; thus, if the emitted photon escapes the cavity before being reabsorbed, its detection places the atom in the superposition $|\psi^\prime\rangle=|g_{-1}\rangle+|g_{+1}\rangle$ (red arrows in Fig.~\ref{levels}a). The atom is now in the ground state with angular momentum perpedicular to the magnetic field, and thus performs Larmor precession.  When subsequently reexcited  by the driven $V$ mode, state
\begin{equation}
|\psi^\prime\rangle=e^{i \phi(\tau)}|e_{-1}\rangle+e^{-i\phi(\tau)}|e_{+1}\rangle
\label{eq:superphase}
\end{equation}
is reached, with phase $\pm\phi(\tau)$ gained through its precession in the ground state. From here the atom can decay back to $|g_0\rangle$, emitting a second $H$-mode photon. The probability to do so depends on $\phi(t)$, giving rise to quantum beats.

In summary, there are two paths for scattering a pair of photons into the $H$ mode (neglect for now the paths to $|g_{\pm2}\rangle$ in Fig.~\ref{levels}a): $|g_0\rangle\rightarrow|e_0\rangle\rightarrow|g_{+1}\rangle\rightarrow|e_{+1}\rangle\rightarrow|g_0\rangle$
and $|g_0\rangle\rightarrow|e_0\rangle\rightarrow|g_{-1}\rangle\rightarrow|e_{-1}\rangle\rightarrow|g_0\rangle$. The phase gained from the ground-state Zeeman shift (precession) differs along the two paths, which interfere to produce oscillations in the rate of delayed coincidences---\textit{i.e.}, in the correlation function $g^{(2)}(\tau)$. Note that after the first photon is detected ``which path'' information is available, since $|g_{+1}\rangle$ and $|g_{-1}\rangle$ are distinguishable in principle. This information is erased by the second detection.




What we have presented is a single-atom idealization. It neglects the presence of more than one atom in the cavity (not admissible for an atomic beam), spontaneous emission to non-cavity modes, the finite cavity decay rate, and the full complement of magnetic sublevels for the employed $F=3$ to $F^\prime=4$ transition. These features are included in a full quantum trajectory treatment, including a Monte-Carlo simulation of an atomic beam \cite{Horvath07}. Two approximations are adopted: (i) the driven mode is treated semiclassically (with absorption still taken into account), and  (ii) reabsorption of $H$-mode photons is neglected. The approximations are justified by our moderate dipole coupling strength.

Consider first an atom prepared in $|g_0\rangle$ that has not yet suffered a spontaneous emission. Let $|a_i\rangle$, $|a^\prime_i\rangle$, and $|a^{\prime\prime}_i\rangle$ denote unnormalized states expanded, respectively, over the $m_i=0$, $m_i=\pm1$, and $m_i=0,\pm2$ sub-manifolds, as indicated by the black, red, and black plus blue levels of Fig.~\ref{levels}a. These states correlate with the scattering of zero, one, and two photons into the $H$ mode. Should the atom undergo a spontaneous emission (to non-cavity modes), the expansion manifolds, after the quantum jump, are unchanged for a $\pi$-emission but move one step to the right or left for a $\sigma$-emission ($m_i\to m_i\pm1$). Keeping track of these shifts, the system state is expanded as
\begin{eqnarray}
|\psi\rangle&=&|0\rangle|A\rangle+|1\rangle\mkern-3mu\left(\sum_{i=1}^N|a^\prime_i\rangle|A\rangle_i\right)\nonumber\\
&&+\sqrt2|2\rangle\sum_{i=1}^N\left(\mkern-2mu\frac12\mkern-2mu\sum_{j\neq i=1}^N|a^\prime_i\rangle|a^\prime_j\rangle|A\rangle_{ij}+|a^{\prime\prime}_i\rangle|A\rangle_i\mkern-2mu\right)\mkern-3mu,\nonumber
\end{eqnarray}
where $|0\rangle$, $|1\rangle$, and $|2\rangle$ denote zero, one, and two photons in the $H$ mode, $N$ is the (time-varying) number of interacting atoms, $|A\rangle=|a_1\rangle|a_2\rangle\ldots|a_N\rangle$, $|A\rangle_i$ is the state $|A\rangle$ with $|a_i\rangle$ omitted from the product, and $|A\rangle_{ij}$ is the state $|A\rangle$ with $|a_i\rangle$ and $|a_j\rangle$ omitted from the product. This base state evolves under the coherent drive and coupling to the $H$ mode, the coming and going of atoms as they transit the cavity, and spontaneous emission. From it, at regular sample times, we initiate the collapsed state $|\psi^\prime\rangle=\hat b|\psi\rangle$, where $\hat b$ annihilates an $H$-mode photon; thus, the ground-state coherence is prepared as an entangled state of $N$ atoms, which then evolves in parallel with $|\psi\rangle$:
\begin{eqnarray}
|\psi^\prime\rangle&=&|0\rangle\sum_{i=1}^{N_0}|b^\prime_i\rangle|A\rangle_i\nonumber\\
&&+|1\rangle\sum_{i=1}^{N_0}\left(\sum_{j\neq i=1}^N|a^\prime_j\rangle|b^\prime_i\rangle|A\rangle_{ij}+|b^{\prime\prime}_i\rangle|A\rangle_i\mkern-2mu\right)\mkern-3mu,
\label{eq:afterstate}
\end{eqnarray}
where $|b^\prime_i\rangle=|a^\prime_i\rangle$ and $|b^{\prime\prime}_i\rangle=2|a^{\prime\prime}_i\rangle$ at the start of a sample, after which $|b_i^\prime\rangle$ ($|b_i^{\prime\prime}\rangle$) and $|a_i^\prime\rangle$ ($2|a_i^{\prime\prime}\rangle$) differ due to their correlation with zero (one) rather than one (two) $H$-mode photons; $N_0\leq N$ is the number of surviving entangled atoms, \textit{i.e.}, those remaining in the cavity.

The source of the quantum beat is the atomic ground-state coherence preserved in the vacuum of the cavity, \textit{i.e.}, the first term in Eq.~(\ref{eq:afterstate}). The \textit{system} ground state---both atoms and cavity---does not decay, and through $\pi$-excitation drives a sustained excited-state oscillation in the atom mirroring Eq.~(\ref{eq:superphase}):
\begin{eqnarray}
\frac{g_{m_i-1}e^{i\delta_g\tau}}{\gamma-i(m_i-1)\Delta}|e_{m_i-1}\rangle+\frac{g_{m_i+1}e^{-i\delta_g\tau}}{\gamma+i(m_i+1)\Delta}|e_{m_i+1}\rangle,
\label{oscillation}
\end{eqnarray}
with $\Delta=\delta_e-\delta_g$, where $\delta_g$ ($\delta_e$) are ground-state (excited-state) Zeeman detunings, $g_{m_i\pm1}$ are Clebsch-Gordon coefficients, and $\gamma$ is the excited state linewidth; $m_i$ tracks the state reached by atom $i$ through possible spontaneous emissions.
From Eq.~(\ref{oscillation}), oscillation at the \textit{ground}-state frequency $\pm\delta_g$ is passed to the probability amplitudes for emitting a second $H$-mode photon through $|b^{\prime\prime}_i\rangle$ and $|a^\prime_j\rangle|b^\prime_i\rangle$ in Eq.~(\ref{eq:afterstate}); note that $|b^\prime_i\rangle$ is a source term driving the equation of motion for $|b^{\prime\prime}_i\rangle$ (the blue wavy lines plus $\pi$-excitation in Fig.~\ref{levels}a). Only the amplitude and phase, but not the frequency of the oscillation, will be affected by a detuning of the drive from the atom.

The probability for emitting a second $H$-mode photon contains a term proportional to $\langle b_i^{\prime\prime}|b_i^{\prime\prime}\rangle$, summed over all surviving entangled atoms. It accounts for the scattering of a first and second photon by the same atom and shows the quantum beat introduced above. There is also a term computed from the norm of $|a_j^\prime\rangle|b_i^\prime\rangle+|a_i^\prime\rangle|b_j^\prime\rangle$, $i\neq j$, which adds probability amplitudes for ``a first photon from atom $i$ and a second from atom $j$,'' and ``a first photon from atom $j$ and a second from atom $i$.'' If the scattered fields were classical, ${\mathcal E}_i^b$ at time $t$ and ${\mathcal E}_i^a$ at time $t+\tau$, one would have the intensity $|{\mathcal E}_j^a{\mathcal E}_i^b+{\mathcal E}_i^a{\mathcal E}_j^b|^2$, where the interference $2{\rm Re}({\mathcal E}_j^{a*}{\mathcal E}_j^b{\mathcal E}_i^{b*}{\mathcal E}_i^a)$ disallows assignment of the first detection (superscript $b$) to the intensity of either source (similarly the second). An assignment may be made in principle, however, since one and only one atom changes its ground state when the photon is detected. The change could be seen if one looked, thus providing ``which-path'' information. As we do not look, the two quantum paths, ``atom $i$ then $j$'' and ``atom $j$ then  $i$'', interfere---after a second detection both atoms have changed state, so the ``which-path'' information is erased. This interference also creates a quantum beat.

A detailed description of the apparatus is given in \cite{norris09a}. The main elements appear in Fig.~\ref{levels}b.  We probe a dilute beam of cold $^{85}$Rb atoms (speed $\sim22\mkern2mu{\rm m/s}$) transiting a $2.2\mkern2mu{\rm mm}$ Fabry-Perot cavity with TEM$_{00}$ mode waist $56\mkern2mu\mu{\rm m}$.  The finesse is 11,000, and the cavity and atomic decay rates, $(\kappa,\gamma)/2\pi=(3.2,6)\times10^6\mkern2mu{\rm s}^{-1}$, are both larger than the maximum dipole coupling, $g/2\pi=1.5\mkern2mu{\rm MHz}$ on the $F=3$, $m=0$ to $F^\prime=4$, $m^\prime=0$ transition; we operate in the intermediate regime of cavity QED, with single-atom cooperativity $C_1=g^2/\gamma\kappa=0.12$ and saturation photon number $n_0=\gamma^{2}/3g^2=5.3$. Splitting of the polarization modes by birefringence is less than $200\mkern2mu{\rm kHz}$.  Atoms enter the cavity optically pumped to the $F=3$, $m=0$ ground state, from which the driven cavity mode excites resonant $\pi$ transitions.

A Glan-Thompson polarizer and zero-order half-wave plate (HWP) placed before the cavity linearly polarize the drive with an extinction ratio that can reach better than $5\times10^{-5}$. A second HWP after the cavity aligns the polarization to a calcite Wollaston prism (PBS) to separate the
$H$- from the $V$-mode.  The former is divided between two avalanche photodiodes at a second beam splitter.  Each detector output goes directly to a time-stamp card, which records a continuous stream of detection times with $164\mkern2mu{\rm ps}$ resolution.  The typical duration of a time series is three minutes or less.

\begin{figure}
\leavevmode \centering
\includegraphics[width=3.1in]{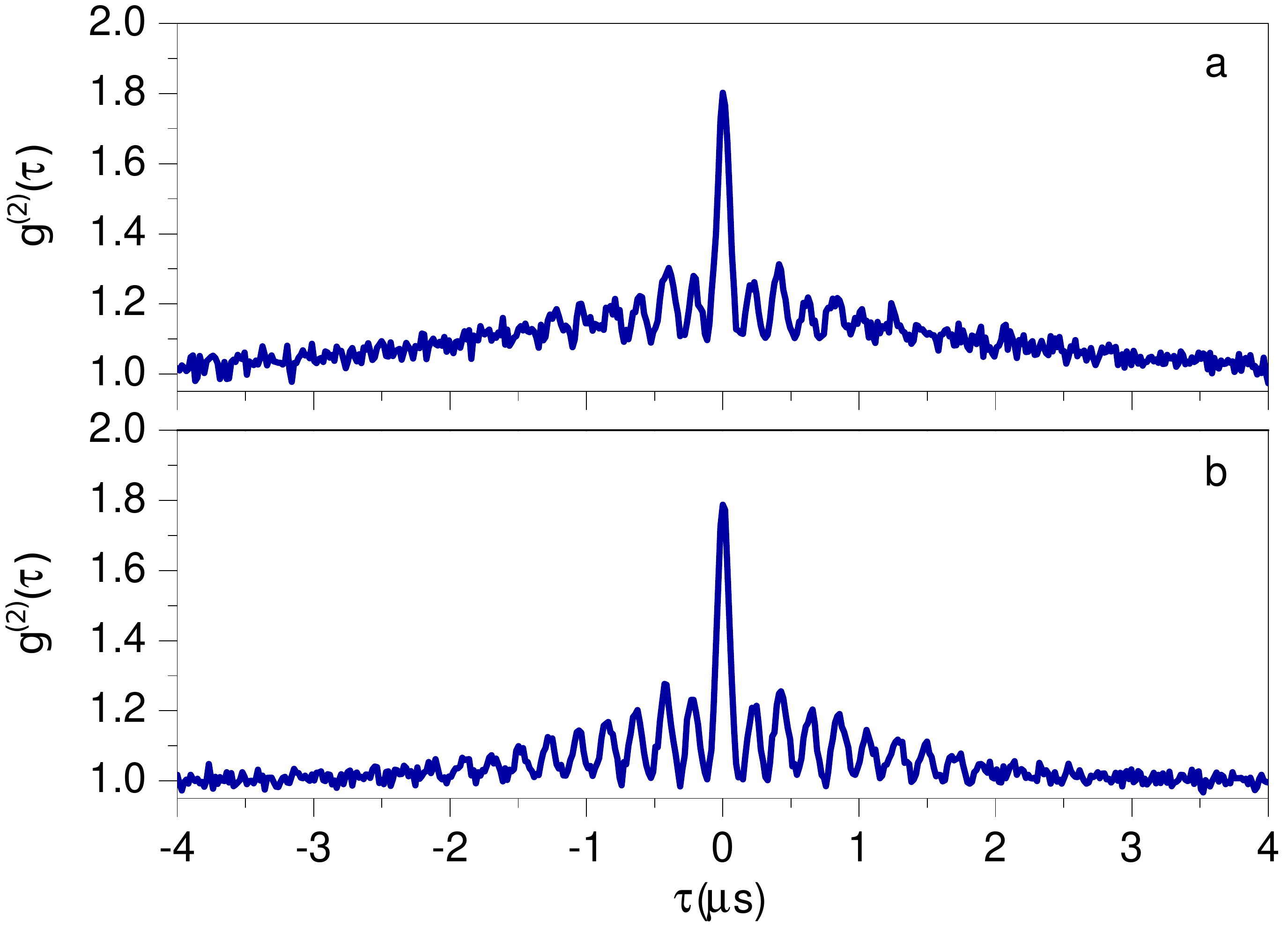}\caption{Intensity correlation function of the $H$ mode for 2--3 photons in the $V$ mode with no atoms present and a $5\mkern2mu{\rm G}$ magnetic field: a) $\bar{N}=0.2$ and b) $\bar{N}=2.0$.  \label{beats}}
\end{figure}

\begin{figure}[t]
\begin{center}
\includegraphics*[width=3.1in]{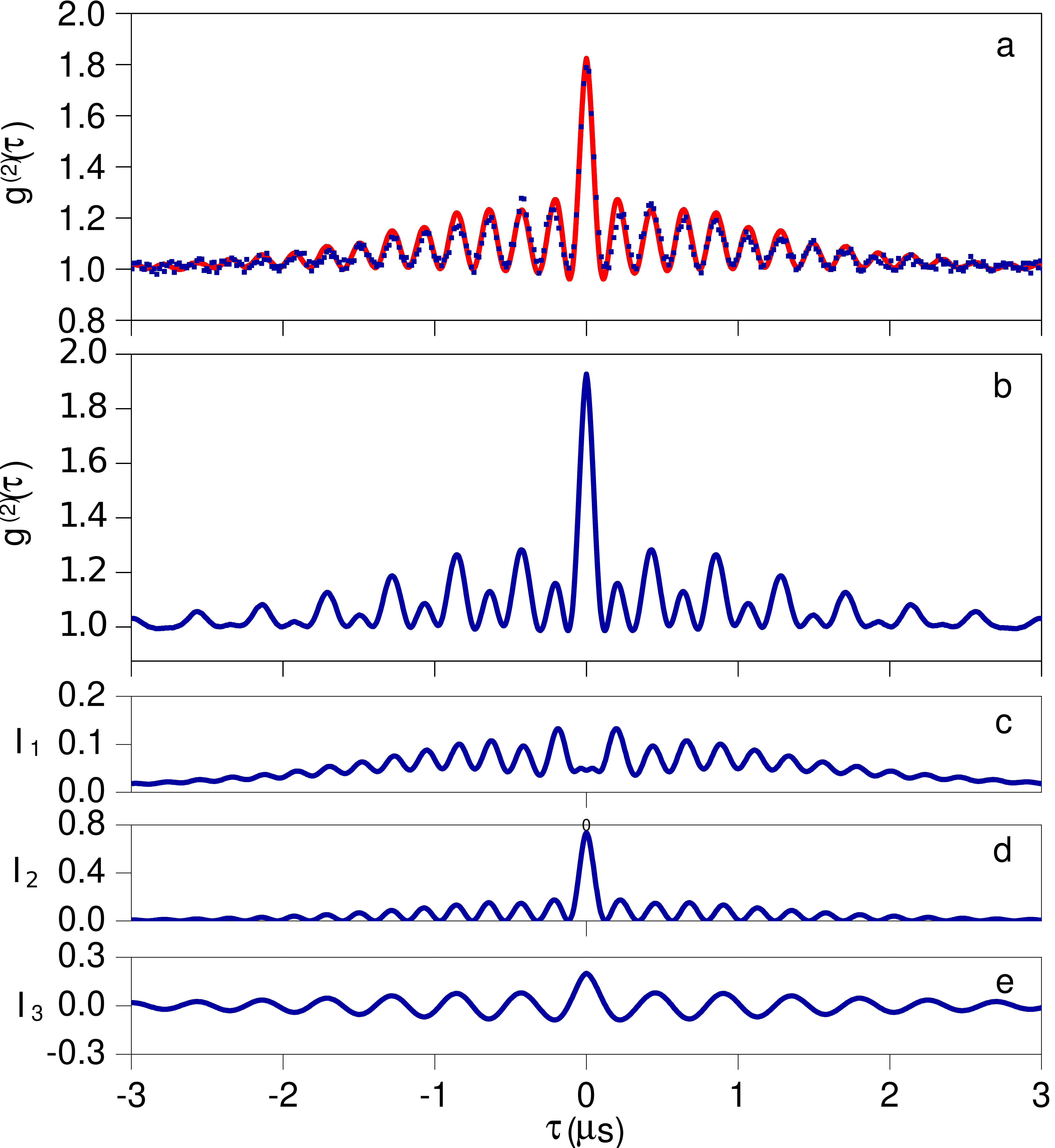}
\caption{\label{theory} Calculated intensity correlation function with the birefringence background set at $1\%$ a) and $10\%$ b) of the $H$-mode photon number; c), d), e) the three interference terms contributing in b); curve a) is plotted against the data of Fig.~\ref{beats} b). Parameters are $\bar N=4$, 4 photons in the $V$ mode with no atoms present, and an atomic beam tilt of $1.3$ degrees.}
\end{center}
\end{figure}

Figure \ref{beats}a displays a measured correlation function at a magnetic field of $5\mkern2mu{\rm G}$, yielding a beat frequency of $4.9\mkern2mu{\rm MHz}$. The $V$ mode is populated with 2--3 photons, on average, when no atoms are present, and the atomic flux corresponds to $\bar N\approx0.2$ effective maximally coupled atoms; most of the time there are no atoms well-coupled to the mode \cite{carmichael99}.
The oscillation has low visibility and sits atop a raised Gaussian background whose correlation time is given by the transit time of an atom ($\approx2.5\mkern2mu\mu{\rm s}$). The dominant beat for small $\bar N$ is that arising from the emission of two photons by the same atom (Fig.~\ref{theory}c). Figure~\ref{beats}b displays the measured correlation function with $\bar N$ larger by a factor of $10$ but otherwise similar conditions.  The beat visibility is improved. The background is also removed, evidence that there is now an equal two-atom quantum beat (Fig.~\ref{theory}d). The beat frequency is reduced to $4.7\mkern2mu{\rm MHz}$, an indirect effect, we believe, of increased absorption by the additional atomic flux, which reduces the $V$-mode photon number---by a factor of two---and thus also the light shifts.  A detailed study of light shifts is planned for separate presentation.


\begin{figure}
\leavevmode \centering
\includegraphics[width=3.1in]{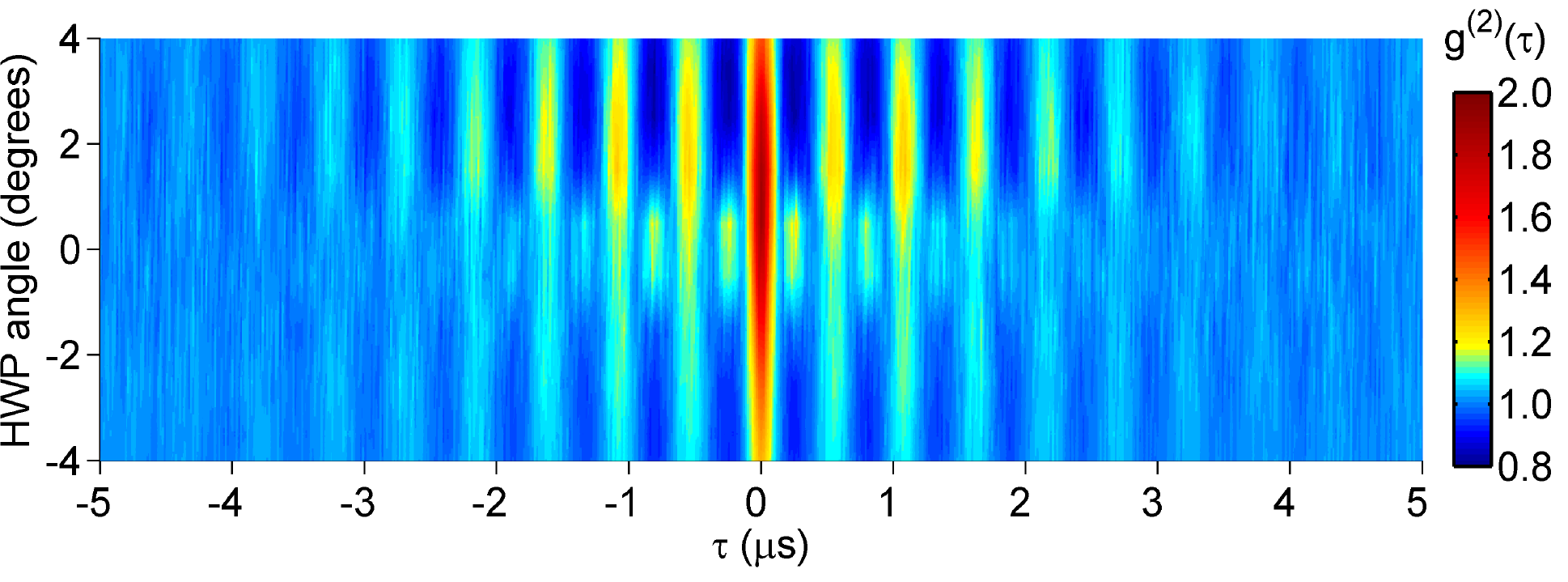}\caption{Evolution of the quantum beat as $V$-mode light is mixed with the $H$ mode.  Upper and lower extremes correspond to approximately six times more detector counts from the $V$ mode than spontaneous emission. The parameters are $0.6$ photons in the $V$ mode, a magnetic field of $4\mkern2mu{\rm G}$, and $\bar{N}=1$.  \label{one-two-omega}}
\end{figure}

Figure \ref{one-two-omega} illustrates the change in the observed beat when the polarization presented to the detector is not taken orthogonal to the polarization of the drive but is allowed to rotate by a few degrees.  The rotation is controlled by changing the angle of the HWP placed between the cavity and the PBS (Fig.~\ref{levels}b).  This mixes a small amount of drive light with the scattered light. With increasing fraction of drive light, the beat is eventually dominated by a homodyne term (Fig.~\ref{theory}e) arising from the correlation of a photon scattered into the $H$ mode with a photon from the drive; thus, as in the two-atom case, interfering time orders yield a quantum beat. This beat oscillates at half the frequency and allows the correlation function to dip below one (\textit{e.g.}, Fig.~\ref{theory}b).  Generally, some drive light is coupled into the $H$ mode through a small birefringence of the cavity mirrors. In Fig.~\ref{one-two-omega}, the evident asymmetry with respect to angle is possibly due to imperfect alignment of the magnetic field with respect to the light polarization.


Quantum trajectory simulations agree well with the measurements. Figure~\ref{theory}a displays a computed correlation function overlaying the data of Fig.~\ref{beats}b.  A mean velocity of $22\mkern2mu{\rm ms}^{-1}$ fits the decay of coherence well. Other parameters, such as the fidelity of the optical pumping, atomic beam tilt, and background from birefringence or elsewhere are more difficult to accurately determine (no background correction is made to the data). Note that with a background light amplitude $\beta$ from birefringence, the post-detection state is $|\psi^\prime\rangle+\beta|\psi\rangle$, and $\beta$ carries noise from the atomic beam and spontaneous emission (absorption on the $V$ mode).  Plausible parameters are used for the plots of Fig.~\ref{theory}, with the quality of the fit primarily determined by the atomic beam density, which controls the relative size of the one- and two-atom quantum beats, and the strength of the drive, which controls the level of spontaneous emission.
For the parameters of Fig.~\ref{theory}a, an atom passing near the cavity axis (within half a mode waist) typically undergoes $\sim10$ spontaneous emissions to non-cavity modes during its transit, yet the coherence, merely passed between different $m_i$-multiplets, is preserved. In contrast, spontaneous emission decoheres an excited-state beat.

Figure~\ref{theory}b displays an example of a correlation function with mixed drive light (Fig.~\ref{one-two-omega}), together with its breakdown into three contributing quantum beats: one-atom interference (frame c), two-atom interference (frame d), and homodyne interference (frame e). The pieces lie in one-to-one correspondence with known terms in the intensity correlation function for a source comprised of many scatterers and a coherent background, \textit{e.g.}, Eq.~(11) of \cite{carmichael78}, where the correlation function is the sum of a single-atom term,  $g^{(2)}_A(\tau)$, two-atom term, $|g^{(1)}_A(\tau)|^2$, and a homodyne term, ${\rm Re[g^{(1)}_A(\tau)]}$.

We have observed ground-state quantum beats in the spontaneous emission from a continuously driven atomic ensemble.  Contrasting deterministic manipulations \cite{wilk07b,weber09}, we demonstrate the spontaneous creation and readout of ground-state coherence, where in the spirit of Forrester \textit{et al.} \cite{forrester55}, we retrieve a hidden beat from the fluctuations.
Our theoretical treatment, which is in good agreement with the measurements, decomposes the signal into a one-atom beat, a two-atom beat (interference of emission time order), and a homodyne beat due to interference with a drive photon mixed through birefringenece. We plan to study a new class of quantum feedback in this system \cite{
barberis10}.



Work supported by NSF, CONACYT, M{\'e}xico, and the Marsden Fund of RSNZ. We thank I. Deutsch for stimulating discussions.
\bibliography{cqed}

\end{document}